\input{aipcheck}

\documentclass[
    ,final            
  ]
  {aipproc}

\usepackage{txfonts}
\layoutstyle{6x9}

\begin{document}

\title{Dynamical masses for the nearest brown dwarf 
binary: $\varepsilon$  Indi Ba,\,Bb\footnote{To appear in the proceedings
of the {\it Cool Stars XV\/} conference (St Andrews, Scotland, July 2008).} 
}


\author{C. V. Cardoso}{
  address={School of Physics, University of Exeter, 
           Stocker Road, Exeter, EX4 4QL, UK}
}

\author{M. J. McCaughrean}{
  address={School of Physics, University of Exeter, 
           Stocker Road, Exeter, EX4 4QL, UK}
}
  
 \author{R. R. King}{
  address={School of Physics, University of Exeter, 
           Stocker Road, Exeter, EX4 4QL, UK}
}

\author{L. M. Close}{
  address={Steward Observatory, University of Arizona, Tucson, USA}
}

\author{R.-D. Scholz}{
  address={Astrophysikalisches Institut Potsdam, Germany}
}

\author{R. Lenzen}{
  address={Max-Planck-Institut f\H{u}r Astronomie, Heidelberg, Germany}
}
  
 \author{W. Brandner}{
  address={Max-Planck-Institut f\H{u}r Astronomie, Heidelberg, Germany}
}

\author{N. Lodieu}{
  address={Instituto de Astrof\'{i}sica de Canarias, Tenerife, Spain}
}

\author{H. Zinnecker}{
  address={Astrophysikalisches Institut Potsdam, Germany}
}

\begin{abstract} We present preliminary astrometric results for the closest 
known brown dwarf binary to the Sun: $\varepsilon$\,Indi\,Ba,\,Bb at a distance 
of 3.626 pc. Via ongoing monitoring of the relative separation of the two 
brown dwarfs (spectral types T1 and T6) with the VLT NACO near-IR adaptive 
optics system since June 2004, we obtain a model-independent dynamical total 
mass for the system of 121 M$_{\rm{Jup}}$, some 60$\%$ larger than the one 
obtained by McCaughrean et al.\ (2004), implying that the system may be as 
old as 5\,Gyr. We have also been monitoring the absolute astrometric motions 
of the system using the VLT FORS2 optical imager since August 2005 to 
determine the individual masses. We predict a periastron passage in early 
2010, by which time the system mass will be constrained to $<$1\,M$_{\rm{Jup}}$ and we will be able to determine the individual masses accurately in a 
dynamical, model-independent manner.  \end{abstract}

\maketitle


\section{Introduction}

The brown dwarf binary $\varepsilon$\,Indi\,Ba,\,Bb \citep{scholz03,mcc04} was 
discovered through a search for high proper motion sources. It shares a high 
common proper motion (4.7"/yr) with the K4V star, $\varepsilon$\,Indi\,A 
(HD\,209100), at a separation of 1459\,AU\@. $\varepsilon$\,Indi\,A has a precise 
HIPPARCOS distance of 3.626$\pm$0.009\,pc \citep{esa97}, making these the 
nearest known brown dwarfs. 
The binary comprises two T dwarfs with spectral types T1 and T6, and the Lyon 
models of Baraffe et al.\ \citep{baraffe03} were used to estimate temperatures 
and masses of 47$\pm$10\,M$_{\rm{Jup}}$ and 28$\pm$7\,M$_{\rm{Jup}}$ for Ba 
and Bb, respectively, assuming an age of 1.3\,Gyr. The age was thought to 
be relatively well constrained at between 0.8 and 2\,Gyr \citep{scholz03}, 
although a significantly older age is favoured by King et al.\ (2008, in
prep; see also \citep{king08}) based on new optical/IR photometry and 
spectroscopy. As a consequence, we are currently pursuing a more accurate 
age estimation for $\varepsilon$\,Indi\,A via asteroseismology. 

However, the brown dwarf masses can be determined independently. The 
separation of the binary at discovery was 0.723", corresponding to 2.65\,AU, 
yielding a nominal orbital period of 16 years assuming the model masses. Thus 
we have been monitoring the relative separation of the two brown
dwarfs using the VLT NACO near-IR adaptive optics system since June 2004 
to obtain a model-independent dynamical total mass for the system. We have 
also been monitoring the absolute astrometric motions of the system using the 
VLT FORS2 optical imager since August 2005 to determine the individual masses. 
Fortunately, the system passed apastron shortly after discovery, making it 
possible to obtain a good preliminary fit to the orbit after only 4 years of 
observations: periastron is predicted to occur in early 2010, at which point 
accurate individual masses will be available, thus providing strong 
constraints on models of these benchmark low-mass objects at an intermediate
age. 

\section{Relative Astrometry}

We have been monitoring the relative orbital motions of $\varepsilon$\,Indi\,Ba 
and Bb with NACO on the VLT in the J, H, and K$_{\rm s}$ bands since 
May 2004: here 
we present a total of 28 epochs obtained up to August 2008 (see Fig.\ 1). 
The NACO N90C10 dichroic is used, diverting 90$\%$ of $\varepsilon$\,Indi\,Ba's 
flux to the NACO IR wavefront sensor: the resulting spatial resolution is 
typically 0.06--0.1" FWHM, meaning simple centroiding can be used to measure 
separations. The wide ($\sim$9") binary system HD\,208371/2 is also measured 
in each band at every epoch to determine the plate-scale and orientation. 
Images are sky-subtracted and flat-fielded, and positions are measured using 
standard IRAF packages. Separations and position angles are calculated and 
errors propagated using a MATLAB script. Relative separations are typically 
accurate to $\pm$2--3\,mas at each epoch, dominated by uncertainties in the 
relative proper motions of HD\,208371/2. Using calibrated position angle and 
separation data for these 28 epochs, we have obtained a preliminary orbital 
fit (see Fig.\ 1) and an estimate of the total dynamical mass of the system 
using the binary package of Gudehus \citep{gudehus01} (see Table 1). Although 
only $\sim$35$\%$ of the orbital period has been covered, the fortunate 
catching of apastron shortly after discovery already leads to a relatively 
accurate fit. Further details of the data reduction and analysis and orbit
fitting will be given by McCaughrean et al.\ (2008, in prep). 

\begin{figure}
\includegraphics[height=.43\textheight]{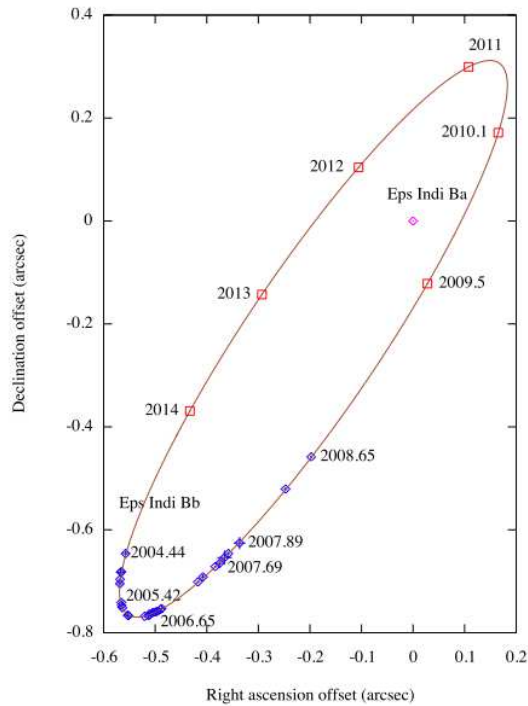}
\caption{Position of $\varepsilon$\,Indi\,Bb relative to Ba over the period 
May 2004 to August 2008. Blue symbols are the data, with blue crosses
showing the errors, while red squares represent the orbital fit prediction. 
Periastron is predicted to be in $\sim$2010.1.}
\end{figure}

\begin{table}
\begin{tabular}{llll}
\hline
\textbf{Period (years)} & 11.2 & $\pm$0.5\\
\textbf{Date of periastron passage} & 2010.1 & $\pm$0.6\\
\textbf{Eccentricity} & 0.53 & $\pm$0.03\\
\textbf{Inclination ($^\circ$)} & $-$77.2 & $\pm$0.7\\
\textbf{Semimajor axis (")} & 0.67& $\pm$0.03\\
\textbf{System mass (M$_\odot$)} & 0.116 & $\pm$0.001 \\
\textbf{System mass (M$_{\rm Jup}$)} & 121 & $\pm$1 \\
   
\hline
\end{tabular}
\caption{System parameters obtained with the preliminary best fit to the 
relative astrometric data using the code of Gudehus \citep{gudehus01}.}
\label{tab:a}
\end{table}

\section{Absolute Astrometry}

We have also been monitoring the absolute motions of $\varepsilon$\,Indi\,Ba 
and Bb across the sky in the I-band with the VLT FORS2 instrument with data 
obtained at 33 epochs from May 2005 to May 2008. Given their proximity to 
the Sun, both brown dwarfs are bright enough to be accurately tracked against 
a net of faint field stars covering a total field of view of 8.53$\times$8.53
arcmin. Images were dark-subtracted, flat-fielded, and sky-subtracted, prior 
to stellar positions being determined in each image using DAOPHOT\@. As the 
FORS2 data are seeing-limited (typically 0.5--0.6" FWHM) and the brown dwarf 
components only partly resolved, PSF fitting must be used to determine their 
positions accurately. A PYRAF script is used to match positions between 
epochs and to determine relative $x,y$ positions throughout. In the field, 
$\sim$\,60 stars are used to construct an internal astrometric reference frame 
and to calculate the relative $x,y$ positions. Of these, 16 have 2MASS 
positions and are used to make the transformation into $\alpha$,$\delta$ 
in the FK5 reference frame. The data clearly show the large (4.7"/yr) 
proper motion of the two sources across the sky, along with a $\sim$0.3" 
parallax wiggle due to their distance of 3.626\,pc (see Fig.\ 2). 
However, the 3 years of data taken to date have been near apastron, where 
the relative orbital motion of $\varepsilon$\,Indi\,Ba and Bb has been rather 
small. While the high acceleration at apastron provides a good preliminary 
estimate of the total system mass, a larger movement on the sky of Bb 
relative to Ba is needed to establish the location of the system barycentre 
and thus determine the mass ratio. This aim should, however, be achieved by 
late 2010 when the system has passed periastron and the maximum relative 
orbital motion has been seen. 

\begin{figure}
\includegraphics[width=\textwidth]{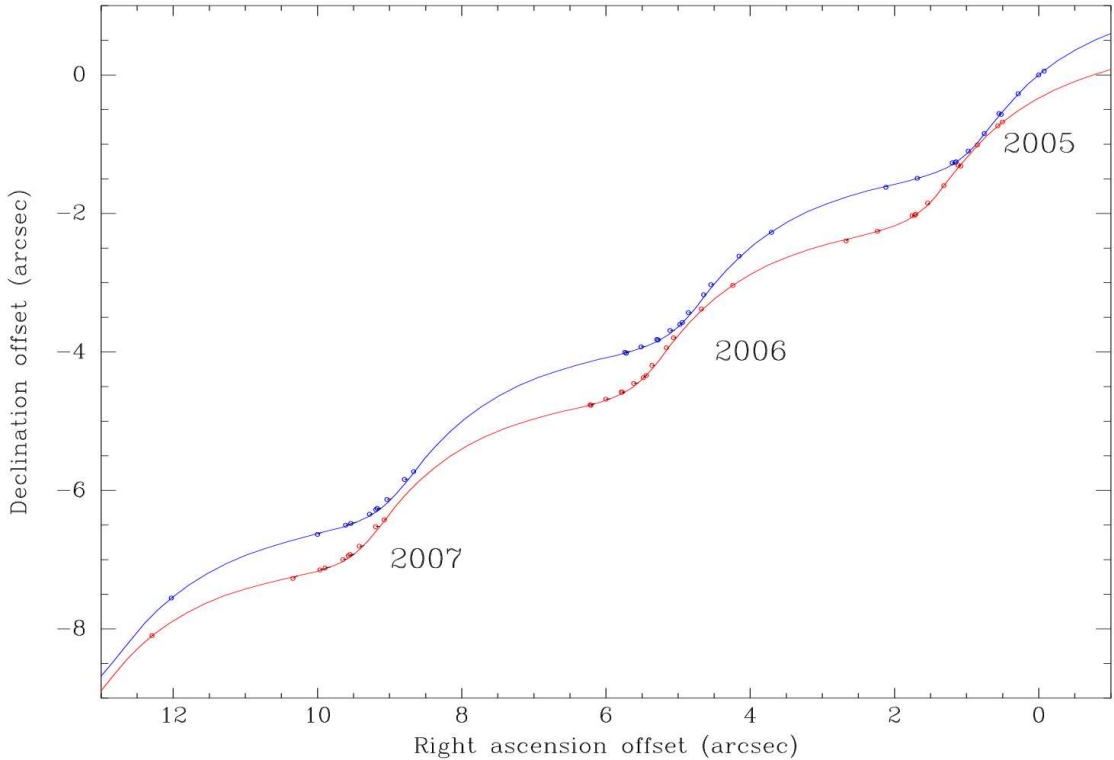}
\caption{Absolute motion of the binary across the field from May 2005 to 
May 2008. The FORS2 data are shown as points while the continuous curves are 
calculated given the relative orbit as fit from the NACO data, the 
$\sim$0.25'' parallax motion due to the 3.626\,pc distance of the source and
its position in ecliptic cordinates, the 4.7'' arcsec/yr proper motion,
and assuming a mass ratio Ba:Bb of 70:50. $\varepsilon$Indi\,Ba is in blue, 
Bb in red. There has been insufficient relative orbital motion as yet to
unambiguously separate the individual masses. 
}
\end{figure}


\section{Results and Future Work}
The key result identified thus far in our study is that the total system 
dynamical mass of 121\,M$_{\rm{Jup}}$ is $\sim$60$\%$ larger than the 
75\,M$_{\rm{Jup}}$ derived by McCaughrean et al.\ \citep{mcc04} based on 
photometry, spectral types, and evolutionary models. However, the dominant 
uncertainty in that latter estimate was due to the age of $\varepsilon$\,Indi\,A, 
taken to be in the range 0.8--2\,Gyr as given by Lachaume et al.\ 
\citep{lachaume99}. If the system were significantly older, the evolutionary 
model masses would need to be larger to account for the observed luminosity 
and thus might be more in line with the directly-measured dynamical masses. 
Indeed, our detailed spectroscopic analysis of the brown dwarfs \citep{king08} 
suggests that the system may in fact be as old as 5\,Gyr, thus potentially 
removing the mass discrepancy. 

Direct dynamical mass determinations for brown dwarfs provide a vital 
calibration of evolutionary and atmospheric models covering a range of 
masses, ages, and metallicities. The first results from such monitoring 
studies are now becoming available, but due to its proximity to the Earth 
and association with a well-known and well-characterised star, the 
$\varepsilon$\,Indi\,Ba,\,Bb binary remains one of the very best suited to 
such studies. The high quality relative orbit data obtained to date clearly 
demonstrate that a very precise system mass can be determined: simulations 
show that with two more years of data including periastron in 2010, the 
system mass will be constrained to significantly better than 
1\,M$_{\rm{Jup}}$, i.e.\
$<$1\%. Furthermore, the unique possibility of following the absolute 
astrometric motions of both sources in wide-field optical imaging means 
that we will also be able to determine the individual masses accurately in 
a dynamical, model-independent manner. 


\begin{theacknowledgments}

This work is funded in part by the EC Sixth Framework Programme Marie Curie 
Research Training Network CONSTELLATION (MRTN-CT-2006-035890).   

\end{theacknowledgments}



\bibliographystyle{aipproc}   
\bibliography{catia1.bib}
\end{document}